\documentclass[a4paper,12pt]{article}
\usepackage[top=60pt,bottom=60pt,left=70pt,right=70pt]{geometry}
\usepackage{amsmath, amssymb, graphicx, hyperref}
\usepackage[greek,english]{babel}
\usepackage{makecell}
\usepackage{authblk}
\usepackage{subcaption}
\usepackage{siunitx}
\usepackage{lineno}
\usepackage{setspace} 

\author[a,*]{Julian Konrad}
\author[a]{Robert Meißner}

\affil[a]{Hamburg University of Technology, Institute of Soft Matter Modeling, Am Irrgarten 3-9, 21073 Hamburg, Germany}
\affil[*]{Corresponding author: \texttt{julian.konrad@tuhh.de}}

\title{Smart Reaction Templating: A Graph-Based Method for Automated Molecular Dynamics Input Generation}
\date{}

\begin{document}

\maketitle
\doublespacing

\section*{Abstract}

Accurately modeling chemical reactions in molecular dynamics simulations requires detailed pre- and post-reaction templates, often created through labor-intensive manual workflows. This work introduces a Python-based algorithm that automates the generation of reaction templates for the \texttt{LAMMPS} REACTION package, leveraging graph-theoretical principles and sub-graph isomorphism techniques. By representing molecular systems as mathematical graphs, the method enables automated identification of conserved molecular domains, reaction sites, and atom mappings, significantly reducing manual effort. The algorithm was validated on three case studies: poly-addition, poly-condensation, and chain polymerization, demonstrating its ability to map conserved regions, identify reaction-initiating atoms, and resolve challenges such as symmetric reactants and indistinguishable atoms. Additionally, the generated templates were optimized for computational efficiency by retaining only essential reactive domains, ensuring scalability and consistency in high-throughput workflows for computational chemistry, materials science, and machine learning applications. Future work will focus on extending the method to mixed organic-inorganic systems, incorporating adaptive scoring mechanisms, and integrating quantum mechanical calculations to enhance its applicability.

\section*{Introduction}

Modeling chemical systems presents a significant challenge due to the intricate interplay of atoms and their interactions. At the core of this complexity lies the necessity to describe chemical reactions, based on the electronic configurations. Quantum mechanics (QM) provides a robust foundation for this task, offering a highly precise framework to capture the behavior of atoms and their electrons. Methods such as ab initio and density functional theory (DFT) have proven to be invaluable tools in this domain \cite{Leach2001, Jensen2016}. However, due to the immense computational demands, these approaches are limited to relatively small systems, typically comprising no more than a few hundred atoms.

To study chemical phenomena at larger spatial and temporal scales, molecular mechanics is an alternative. Simply speaking, this method simplifies atomic interactions into a "bead-and-spring" model, where intra-molecular interactions are approximated as simple linear springs, and non-bonded interactions are represented using potential functions \cite{Karplus2002}. While less detailed than quantum mechanical methods, molecular mechanics enables the efficient simulation of systems containing thousands or even millions of atoms, facilitating insights into macroscopic properties and dynamic behavior.

One downside of such molecular mechanics (MM) models, which rely on fixed parameters for bonds, angles, dihedrals, and charges is that they are typically unsuitable for capturing chemical reactivity due to their static nature. To overcome this limitation, approaches such as the charge equilibration method \cite{Rappe1991} and reactive force fields have been developed. These methods enable dynamic charge adjustments and the simulation of bond formation and breaking on the basis of some smooth bond-order dependent functions, thereby allowing for modeling of chemical reactions. ReaxFF, in particular, is a versatile tool capable of handling a wide range of reactive systems, but it often requires substantial computational resources and careful parametrization to align with experimental data or high-level quantum mechanical calculations \cite{vanDuin2001, Chenoweth2008}. Similarly, tailor-made reactive force fields offer precise solutions for specific systems but demand labor-intensive calibration processes to ensure their accuracy and reliability \cite{Konrad2021}.

Another approach that requires less parametrization from quantum mechanical calculations is the Smooth Topology Transfer method, as described by Meißner \textit{et al.}~ \cite{Meiner2020}. This method enables a seamless transition between the topologies of reactants and products by blending molecular forces and energies in a controlled manner. While this approach reduces the need for extensive quantum mechanical data and incorporates only QM energy correction terms, defining the reactant and product topologies remains a labor-intensive process.

The REACTION package in \texttt{LAMMPS} \cite{Thompson2022} provides functionality for simulating chemical reactions in molecular systems by utilizing pre- and post-reaction templates to define the topologies of reactants and products \cite{Gissinger2017, Gissinger2020, Gissinger2024}. The process involves mapping individual atoms between these templates and defining reaction criteria, such as distance thresholds or specific bonding conditions. While fix bond/react facilitates the modeling of reactions in large-scale simulations, preparing accurate templates remains a significant challenge. These templates must account for bonds, angles, dihedrals, constraints and special neighbor rules to ensure the consistency of the system’s topology.

To address the challenges of template preparation, \texttt{AutoMapper} was developed as a Python-based tool to automate much of the workflow for fix bond/react simulations in \texttt{LAMMPS} \cite{Bone2022}. \texttt{AutoMapper} significantly reduces the manual effort required by generating reaction templates from \texttt{LAMMPS} input files. However, users must still manually specify which bonds are formed or broken during the reaction. 

A recent study introduced the fully automated algorithm \texttt{PX-MDsim} for the cross-linking of polyamides \cite{Peng2025}. This platform, based on the PXLink framework, automates the entire cross-linking simulation process - from input preparation and initial system construction to force field generation, functional group identification, and charge distribution updates. However, its applicability is limited to monomers containing amino and carboxyl groups.

To enable high-throughput screening of materials and generate diverse structural datasets for machine learning applications, it is essential to fully automate the mapping process between reactants and products. More importantly, our approach imposes no restrictions on specific monomers or types of chemical reactions, making it broadly applicable. In this work, we present a Python-based algorithm that relies exclusively on the definition of \texttt{LAMMPS} data files. These files can be conveniently generated from simple SMILES \cite{Weininger1988} strings using tools such as \texttt{LigParGen} \cite{Dodda2017}, \texttt{Open Babel} \cite{OBoyle2011}, and \texttt{Moltemplate} \cite{Jewett2021}.

By representing molecules as graphs, the algorithm employs graph isomorphism techniques, optimization, and neighborhood analysis to automatically map reactant and product atoms and identify reaction sites. This approach eliminates much of the manual effort traditionally associated with preparing reaction templates and ensures consistent and scalable workflows for complex simulations and data generation tasks.

\section*{Methods}

Molecular structures are represented as graphs \(G = (V, E)\), where \(V\) is the set of nodes (atoms) and \(E\) is the set of edges (bonds). Each node \(v \in V\) is assigned attributes such as atomic mass \(m(v)\), atom type \(t(v)\), atomic charge \(q(v)\), and a component identifier \(c(v)\), which specifies the molecule to which the atom belongs. Edges \(e = (v, u) \in E\) connect nodes to represent bonds determined by the bonding information in the input data.

This graph-based representation facilitates the systematic identification of molecular features such as bonds (direct connections), angles (triplets of connected nodes), and higher-order interactions like dihedrals. These features are derived by iteratively traversing neighboring nodes and grouping them into structural patterns. Each identified feature is verified for uniqueness using the atom type \(t(v)\) of the participating nodes, ensuring consistent classification across all molecules. 

For chemical reactions, the reactants and products are represented as unified graphs, \(G_\text{reac} = (V_\text{reac}, E_\text{reac})\) and \(G_\text{prod} = (V_\text{prod}, E_\text{prod})\), respectively. These graphs are constructed by combining the individual molecular graphs of the reactants \(G_\text{r}\) and products \(G_\text{p}\):
\[
V_\text{reac} = \bigcup_{i=1}^{n_\text{reac}} V_{\text{r},i}, \quad 
E_\text{reac} = \bigcup_{i=1}^{n_\text{reac}} E_{\text{r},i},
\]
\[
V_\text{prod} = \bigcup_{j=1}^{n_\text{prod}} V_{\text{p},j}, \quad 
E_\text{prod} = \bigcup_{j=1}^{n_\text{prod}} E_{\text{p},j},
\]
where \(n_\text{reac}\) and \(n_\text{prod}\) denote the number of reactant and product molecules, and \(G_{\text{r},i} = (V_{\text{r},i}, E_{\text{r},i})\) and \(G_{\text{p},j} = (V_{\text{p},j}, E_{\text{p},j})\) are the graphs of individual reactant and product molecules.
The construction of \(G_{\text{r},i}\) and \(G_{\text{p},j}\) involves parsing input data to extract atomic properties \(m(v)\), \(t(v)\), \(q(v)\), and \(c(v)\) for each atom. Based on this information, edges \(E_{\text{r},i}\) or \(E_{\text{p},j}\) are established according to bonding information. To ensure consistency and avoid conflicts, globally unique node identifiers are assigned when combining individual graphs into \(G_\text{reac}\) and \(G_\text{prod}\). The component identifier \(c(v)\) ensures that nodes remain associated with their respective molecules during the mapping process:
\[
    c(v) = k, \quad \text{if } v \in V_{\text{r},k} \text{ or } v \in V_{\text{p},k},
\]
where \(k\) corresponds to the index of the reactant or product molecule.

This unified representation of reactants and products as \(G_\text{reac}\) and \(G_\text{prod}\) allows for efficient computation of sub-graph alignments, which are essential for identifying reaction sites and transformations. By retaining the modularity of individual molecular graphs, the method ensures scalability and accuracy in handling complex chemical systems.    
    
The identification of structurally conserved regions between the reactants and products is achieved through sub-graph isomorphism detection. This step determines correspondences between parts of the reactant and product molecular graphs that are not directly involved in the chemical reaction.
    
A sub-graph isomorphism is defined as a bijection \( f: V_\text{r}' \to V_\text{p}' \), where \( V_\text{r}' \subseteq V_\text{reac} \) and \( V_\text{p}' \subseteq V_\text{prod} \), such that:
\[
    (v, u) \in E_\text{reac} \implies \big(f(v), f(u)\big) \in E_\text{prod},
\]
and node attributes such as atomic mass and type are preserved:
\[
    m(v) = m(f(v)), \quad t(v) = t(f(v)), \quad \forall v \in V_\text{r}'.
\]

It might be necessary to perform multiple iterations of sub-graph isomorphism search. Since the reaction can occur at the center of a reactant, it results in two conserved regions that may not be fully captured in a single iteration, as they are no longer connected but are separated by the reaction site.

In each iteration, the largest conserved sub-graph is identified. For each reactant molecule and product molecule, the unmapped portions of the reactant and product graphs are extracted. These sub-graphs exclude all nodes and edges already matched in previous iterations. Initially, these sub-graphs exclude no nodes or edges, as the exclusion only occurs after the mapping. Among the unmapped portions, a mapping is sought to maximize the size of the conserved region:
\[
    f^* = \arg\max_{f} |V_\text{r}' \cap f(V_\text{p}')|,
\]
where \( f \) represents all possible bijections preserving node attributes and connectivity. The result of this step is a mapping of the largest structurally conserved sub-graph between the reactant and product graphs. The remaining unmapped portions of the graphs are then processed in subsequent iterations to identify smaller conserved sub-graphs.
    
The mappings \(f^*\) from all iterations are combined to form the overall correspondence between the reactant and product graphs in conserved regions. This ensures that each node in the reactant graph is mapped to at most one node in the product graph and that the mapping encompasses all conserved regions, regardless of their size or separation. This approach guarantees completeness by systematically processing both large and small conserved regions. Two iterations are sufficient to ensure that conserved regions on both sides of the reaction site are fully mapped and that fragmented components, which might otherwise be missed, are included in the final mapping of the conserved regions. 

After identifying conserved substructures using sub-graph isomorphism, the remaining unmapped nodes, which correspond to the reaction site, are aligned based on a similarity score that incorporates structural and chemical properties, as well as the local molecular environment.

The similarity score \( S \) for a pair of unmapped nodes \( \tilde{v}_\text{r} \in \tilde{V}_\text{reac} \) and \( \tilde{v}_\text{p} \in \tilde{V}_\text{prod} \) is computed iteratively as:
\[
S(\tilde{v}_\text{r}, \tilde{v}_\text{p}) = \frac{1}{|V^{\ast}|} \sum_{k=1}^{d} \bigg( \alpha \cdot \delta_t^{(k)} + \beta \cdot \delta_m^{(k)} + \gamma \cdot (N_p^{(k)} + N_m^{(k)}) \bigg),
\]
where \( k \) represents the current neighborhood depth in the graph, with \( d \) being the smaller of the longest shortest paths from \( \tilde{v}_\text{r} \) to any node in \( V_{\text{r}} \) or \( \tilde{v}_\text{p} \) to any node in \( V_{\text{p}} \), ensuring a consistent maximum depth in both graphs. The weights \( \alpha \), \( \beta \), and \( \gamma \) control the relative importance of mass, type, and neighborhood similarity in the score. The terms \( \delta_t^{(k)} \) and \( \delta_m^{(k)} \) are indicator functions that are \( 1 \) if the atom types and masses of the nodes \( \tilde{v}_\text{r} \) and \( \tilde{v}_\text{p} \) are identical at depth \( k \), and \( 0 \) otherwise. The terms \( N_p^{(k)} \) and \( N_m^{(k)} \) measure the neighborhood similarity in terms of atom type and mass, respectively, and are computed as:
\[
N_p^{(k)} = \sum_{j} \delta_t^{(k)}(j), \quad N_m^{(k)} = \sum_{j} \delta_m^{(k)}(j),
\]
where \( j \) indexes the neighboring nodes of \( \tilde{v}_\text{r} \) and \( \tilde{v}_\text{p} \) at depth \( k \). The normalization factor \( |V^{\ast}| \) represents the number of nodes considered in the iterative process and is given by:
\[
|V^{\ast}| = \max(|V_\text{reac}^{(k)}|, |V_\text{prod}^{(k)}|).
\]
Once the similarity scores are computed, the cost matrix \( \mathbf{C} \) is constructed as:
\[
C_{ij} = -S(\tilde{v}_\text{r}^i, \tilde{v}_\text{p}^j),
\]
where each entry represents the negative similarity score between nodes \( \tilde{v}_\text{r}^i \) and \( \tilde{v}_\text{p}^j \). To ensure numerical stability, the matrix is normalized. Finally, the optimal one-to-one mapping is determined by solving the linear sum assignment problem:
\[
f^* = \arg\min_{f} \sum_{(i, j) \in f} C_{ij},
\]
where \( f \) represents the optimal node correspondence between \( \tilde{V}_\text{reac} \) and \( \tilde{V}_\text{prod} \), and the summation runs only over the assigned pairs.

The weights \(\alpha\), \(\beta\), and \(\gamma\) are normalized such that their sum equals 1, ensuring a balanced contribution of mass, type, and neighborhood similarity. 

If a molecule is composed of structurally identical paths, certain sub-graph traversals become indistinguishable. Thus, isomorphic node anchors \( p_1\) serve as fixed reference points in the mapping process. Iterative adjustments are made to maximize structural consistency by swapping nodes and re-evaluating their associations based on bond connectivity. Subsequently, paths originating from the reassigned nodes are evaluated to ensure connectivity consistency. For a path $\{p_1, p_2, \dots, p_k\}$ originating from a node \(p_1\), the reassignment is refined iteratively to maximize structural alignment:
\[
\mathcal{A}_{\text{opt}} = \arg\max_{f} \sum_{k} S^*(p_k, f(p_k)),
\]
where \(f\) represents the node mapping, and \(S^*\) is a similarity function based on node attributes and connectivity.
This process guarantees a complete and accurate mapping, ensuring alignment consistency across all conserved regions and reaction-site nodes. \\

To ensure consistent alignment between nodes in the reactant and product graphs, discrepancies in neighborhood connectivity are addressed when \( G_{\text{reac}} \) contains more nodes than \( G_{\text{prod}} \). The adjustment process is as follows: For each pair of aligned nodes \((v_\text{r}, v_\text{p})\) the sizes of their respective neighborhoods are compared. Let \(\mathcal{N}(v_\text{r})\) and \(\mathcal{N}(v_\text{p})\) denote the neighborhoods of \(v_\text{r}\) and \(v_\text{p}\), respectively. A mismatch is identified if:
\[
|\mathcal{N}(v_\text{r})| \neq |\mathcal{N}(v_\text{p})|.
\]
To resolve these mismatches, hydrogen-like nodes are swapped. The algorithm identifies candidates for swapping by iterating over the neighbors of $v_r$ and $v_\text{p}$. The sets of such nodes are denoted as:
\[
\mathcal{N}_\text{swap}(v_\text{r}) = \{v_\text{r}' \in \mathcal{N}(v_\text{r}) \mid m(v_\text{r}') \approx 1.008\}, \quad 
\mathcal{N}_\text{swap}(v_\text{p}) = \{v_\text{p}' \in \mathcal{N}(v_\text{p}) \mid m(v_\text{p}') \approx 1.008\},
\]
where \(m(x)\) represents the mass of node \(x\).
Pairs of nodes \((v_\text{r}', v_\text{p}')\) from these sets are identified for swapping, ensuring that connectivity remains consistent. 

Reaction sites are identified by analyzing changes in bond connectivity between aligned nodes. For each aligned pair \( (v_\text{r}, v_\text{p}) \), the bond differences are quantified using \( \Delta E(v_\text{r}, v_\text{p}) \):
\[
\Delta E(v_\text{r}, v_\text{p}) = \big( N_p(v_\text{p}) \setminus N_r(v_\text{r}) \big) \cup \big( N_r(v_\text{r}) \setminus N_p(v_\text{p}) \big).
\]
To determine the key reaction site, the structural importance of each node is assessed using changes in eigenvector centrality \( \Delta C \), where centrality \( C \) is defined as:
\[
C(v) = \frac{1}{\lambda} \sum_{u \in N(v)} C(u),
\]
with \( \lambda \) being the largest eigenvalue of the adjacency matrix \cite{Bonacich2001}. The reaction sites is ultimately identified by selecting nodes associated with the largest combined changes in \( \Delta E \) and \( \Delta C \), ensuring that both bond rearrangement and network influence are considered.

The algorithm produces three primary outputs, tailored for efficient use in simulations. First, it generates reactant and product templates that specify the atomic structures, bonds, angles, and other interactions for \( G_\text{reac} \) and \( G_\text{prod} \). To minimize template size, only the reaction site and its neighboring atoms, determined by a distance cutoff of four in the graph representation, are included. This reduction is achieved by calculating the shortest path distances from the reaction site nodes and selecting atoms within the defined cutoff radius. Edges corresponding to these nodes are also identified, ensuring the inclusion of all relevant structural and interaction information. Second, the algorithm produces mapping, identifies the reaction sites, and accounts for any atoms that are created or deleted. 

The applicability of the algorithm was exemplified by investigating a selection of polymer classes and chemical reactions. Specifically, the algorithm was applied to the poly-addition of butanediol (BD) and methylenediphenylisocyanate (MDI), the poly-condensation of cyclohexane-1,3,5-tricarboxylic acid (HT) and 1,3 phenylenediamine (MPD) \cite{Arthur1989}, and a alchemical type of chain-polymerization between two 1-buthene molecules to octane. The molecular SMILES strings and force fields used are summarized in Table~\ref{tab:reactions}.

\begin{table}[h!]
\centering
\caption{The SMILES strings for the reactions under study are provided, reflecting the reactants (r) and the product (p). The \texttt{LAMMPS} data files, derived from these strings, were parameterized using the corresponding force field.}
\renewcommand{\arraystretch}{2} 
\begin{tabular}{|c|l|c|}
\hline
Reaction (FF) & MOL & SMILES  \\ 
\hline
\makecell{Poly-addition \\ (OPLS-AA)} & \makecell[{{l}}]{ $\text{r}_1$ (MDI) \\ $\text{r}_2$ (BD) \\ $\text{p}_1$  }    & \makecell[{{l}}]{O=C=Nc2ccc(Cc1ccc(N=C=O)cc1)cc2 \\ C(CCO)CO \\ O=C=Nc2ccc(Cc1ccc(NC(=O)OCCCCO)cc1)cc2}  \\ 
\hline
\makecell{Poly-condensation \\ (GAFF) } & \makecell[{{l}}]{ $\text{r}_1$ (HT) \\ $\text{r}_2$ (MPD) \\ $\text{p}_1$ } & \makecell[{{l}}]{O=C(O)C1CC(C(=O)O)CC(C(=O)O)C1 \\ Nc1cccc(N)c1 \\ Nc2cccc(NC(=O)C1CC(C(=O)O)CC(C(=O)O)C1)c2}  \\ 
\hline
\makecell{Chain-polymerization \\ (GAFF) } & \makecell[{{l}}]{ $\text{r}_1$  \\ $\text{r}_2$ \\ $\text{p}_1$ } & \makecell[{{l}}]{C=CCC \\ C=CCC \\ CCCCCCCC}  \\ 
\hline
\end{tabular}
\label{tab:reactions}
\end{table}

Based on these inputs, the initial \texttt{LAMMPS} data files were generated applying the desired force fields. For the OPLS-AA force field \cite{Jorgensen1996}, LigParGen \cite{Dodda2017} was employed, while GAFF (General Amber Force Field) \cite{Wang2004} parameters were generated using OpenBabel \cite{OBoyle2011} in combination with the Antechamber module of AmberTools \cite{Case2023}. To achieve accurate charges, a RESP fit was performed following the parametrization process \cite{Woods2000}.

Subsequently, the algorithm was used to generate the reactant and product templates, along with the necessary mapping files for the simulations. A full simulation, including crosslinking to bulk polymer structure, was performed for the poly-condensation between HT and MPD, using \texttt{LAMMPS}. 240 HT molecules and 320 MPD molecules, comprising in total 10233 atoms, were equilibrated in a simulation box for 3\,\si{ns} at room temperature, followed by heating to the cross-linking temperature of 600\,\si{K} for 2\,\si{ns}. Crosslinking was then performed for 5\,\si{ns} using the REACTION package in \texttt{LAMMPS} \cite{Gissinger2017}, with NVE/limit applied for 500 \si{fs}, restricting the displacement of reaction sites to 0.0015 {\AA} per \si{fs}. During the simulation, the hydroxyl groups of the carboxylic acids were constrained using the SHAKE algorithm to enhance stability \cite{Ryckaert1977}. Constant temperature and pressure were maintained using the Nosé-Hoover thermostat and barostat with a damping time \(\tau\) of 100\,\si{fs} and a pressure damping time \(\tau\) of 1000\,\si{fs} at 1\,\si{atm} \cite{Evans1985}. Periodic boundary conditions were applied with a time step of 0.5\,\si{fs}. Lennard-Jones interactions and electrostatics were calculated using a cutoff of 12\,{\AA}, while the long-range electrostatic interactions were treated using the particle-particle particle-mesh (PPPM) method with a $k$-space accuracy of 0.0004\,\text{kcal/mol\,$\cdot$\,\AA} \cite{Hockney1988}.

\section*{Results \& Discussions}

The weighting parameters were empirically determined to provide an optimal balance for capturing a diverse range of reaction types. The initial parameter values were set to \(\alpha = 0.5\), \(\beta = 0.25\), and \(\gamma = 0.25\).

\begin{figure}[htb!]
    \centering
    \begin{minipage}{0.33\textwidth}
        \centering
        \includegraphics[width=1.0\textwidth]{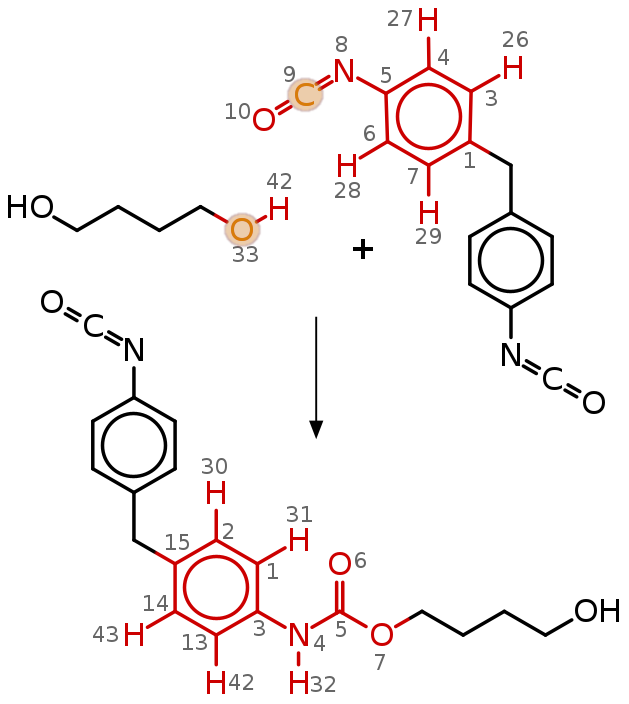}
    \end{minipage} \hfill
    \begin{minipage}{0.63\textwidth}
        \centering
        \includegraphics[width=1.0\textwidth]{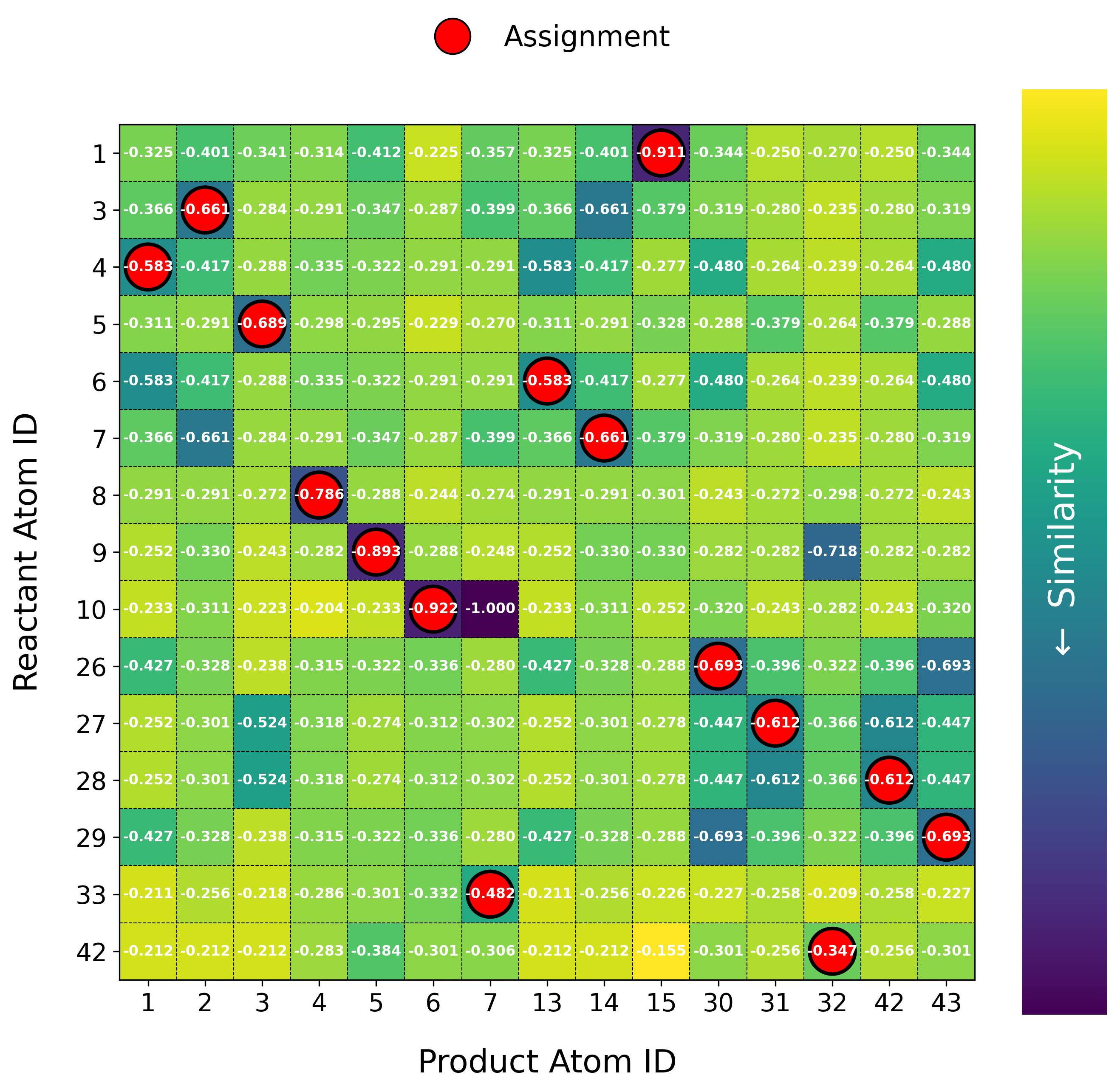}
    \end{minipage}
    \caption{The mapping process for the poly-addition of BD and MDI (left) is depicted. The chemical structures of the reactants and product distinguish between conserved domains (black) and the nodes that were mapped based on the similarity criterion (red). The atom labels correspond to the nodes of the similarity matrix on the right. Final assignments are shown in red.}
    \label{fig:polyaddition}
\end{figure}

The versatility of the algorithm is demonstrated through the poly-addition reaction between butanediol (BD) and methylenediphenylisocyanate (MDI). In the initial step, sub-graph isomorphism successfully identified conserved molecular domains, mapping 30 out of 45 nodes, as shown in black in Figure~\ref{fig:polyaddition}. The conserved domains are notably smaller than expected since the aromatic ring not being directly involved in the reaction. However, the force field’s atom type definition distinguishes between isocyanate and urethane substitutions, leading to the assignment of the aromatic ring to the reactive domain. Consequently, the definition of conserved molecular regions depends on the provided input data and force field, emphasizing the algorithm's precision in isolating chemically invariant domains.

The subsequent alignment, based on similarity scores, accurately mapped nodes corresponding to heavy atoms (C, N, and O) directly involved in the reaction, as highlighted in red in Figure~\ref{fig:polyaddition}. The expected conserved domain of the aromatic ring was correctly assigned, demonstrating the robustness of the algorithm even with more complex force fields definitions and large reactive domains that seemingly contain conserved regions. Furthermore the algorithm correctly matches indistinguishable paths in the graph structure. For instance, the para-phenyl substitution in the product allows two paths from atom ID 3 to 15, either through atoms 1 and 2 or through atoms 13 and 14. The algorithm also successfully identified the proton transfer from the hydroxyl group of BD to the urethane group of the product. Reaction-initiating atoms, including the reactive sites of the hydroxyl and isocyanate groups, were automatically detected and are highlighted in orange. To enhance the efficiency of subsequent simulations, the final \texttt{LAMMPS} templates were stripped to include only 31 atoms, representing the essential reactive domains and their immediate environment. These optimized templates are provided in the supplementary materials. This example highlights the algorithm's adaptability in accurately handling cases involving atom rearrangements into newly formed chemical groups and indistinguishable paths, ensuring precise node alignments and robust mapping in complex reaction systems. \\

\begin{figure}[htb!]
    \centering
    \begin{minipage}{0.48\textwidth}
        \centering
        \includegraphics[width=1.0\textwidth]{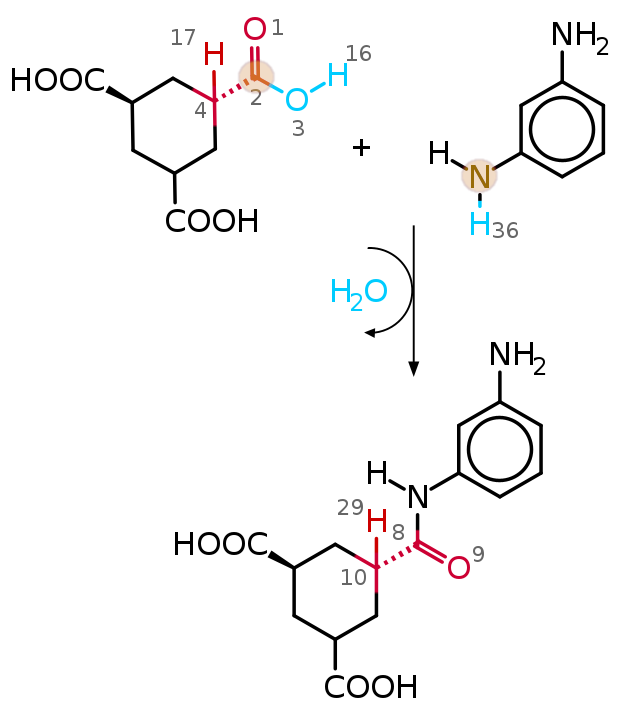}
    \end{minipage} \hfill
    \begin{minipage}{0.48\textwidth}
        \centering
        \includegraphics[width=1.0\textwidth]{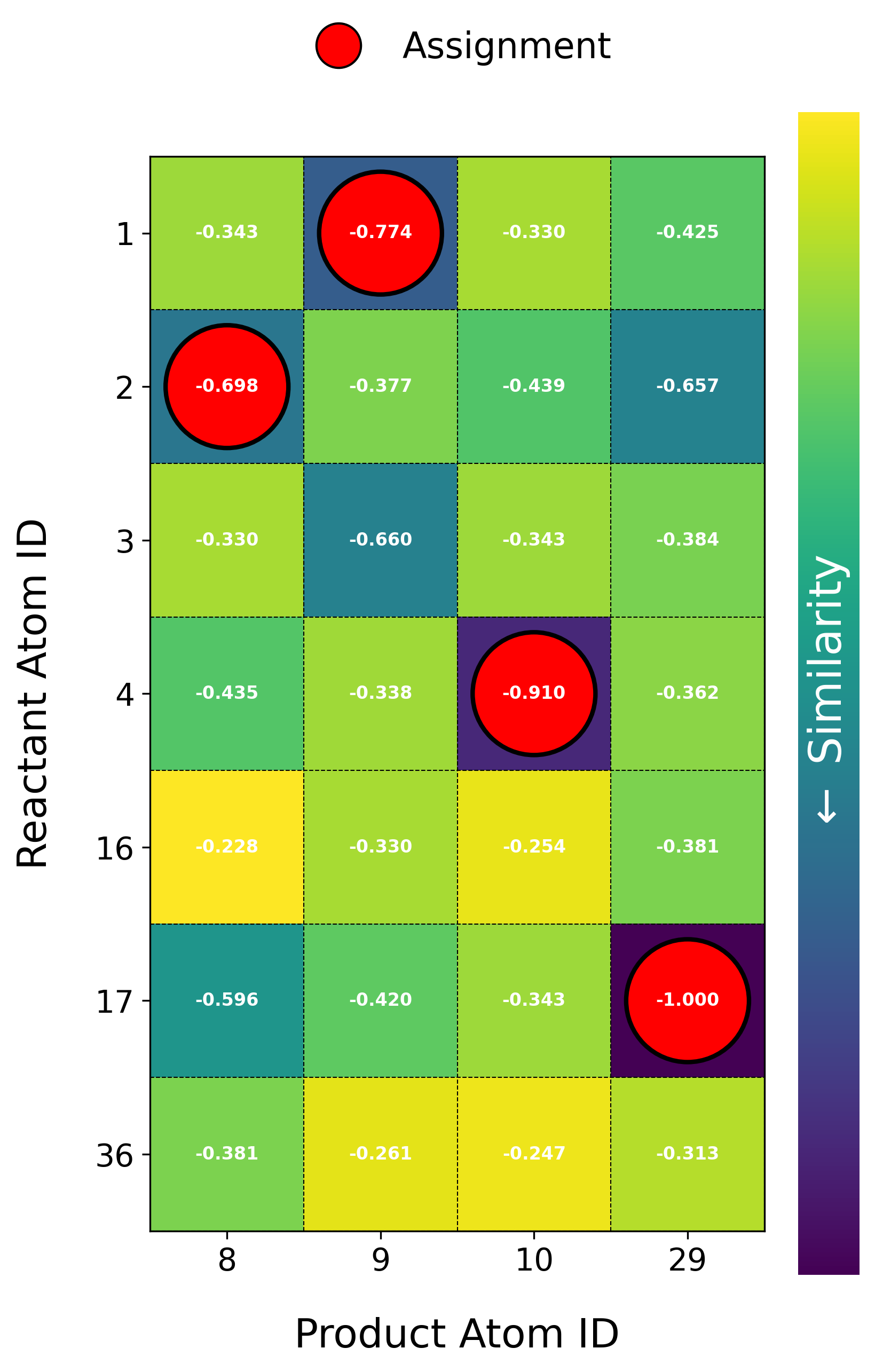}
    \end{minipage}
    \caption{The representation of the poly-condensation reaction between HT and MDP (left) shows conserved domains in black, mapped similar atoms in red, and atoms that are eliminated in blue. Automatic detection of the reaction-initiating atoms is highlighted in orange. The atom indices correspond to the node numbers of the graph representation of the molecules. The calculated similarity scores (right) for the nodes from \(G_\text{reac}\) and \(G_\text{prod}\), along with the assignment via linear sum assignment, illustrate the convergence of the algorithm.}
    \label{fig:polycondensation}
\end{figure}

In Figure~\ref{fig:polycondensation}, the reactants and products for the poly-condensation reaction between HT and MDP are illustrated. Conserved molecular structures, identified through sub-graph isomorphisms are highlighted in black with 36 of 40 nodes. The remaining atoms participating in the chemical reaction, are labeled with their respective indices from the graphs \(G_\text{reac}\) and \(G_\text{prod}\).

The similarity scores for the unmapped nodes were calculated, and the algorithm achieved an optimal assignment, as highlighted in red in Figure~\ref{fig:polycondensation}, accurately mapping the acidic group to the newly formed amide bond. The water elimination process was identified and is represented by blue atoms. Reaction-initiating atoms were automatically detected and highlighted in orange. The final reaction template, consisting of 27 atoms, was derived by minimizing the structural representation to include only the reaction site and its immediate environment, reducing computational complexity while preserving chemical accuracy. The templates are provided in the supplementary materials. Figure~\ref{fig:system} shows the corresponding reactant and product templates.

\begin{figure}[htb!]
    \centering
    \begin{minipage}{0.38\textwidth}
        \centering
        \includegraphics[width=1.0\textwidth]{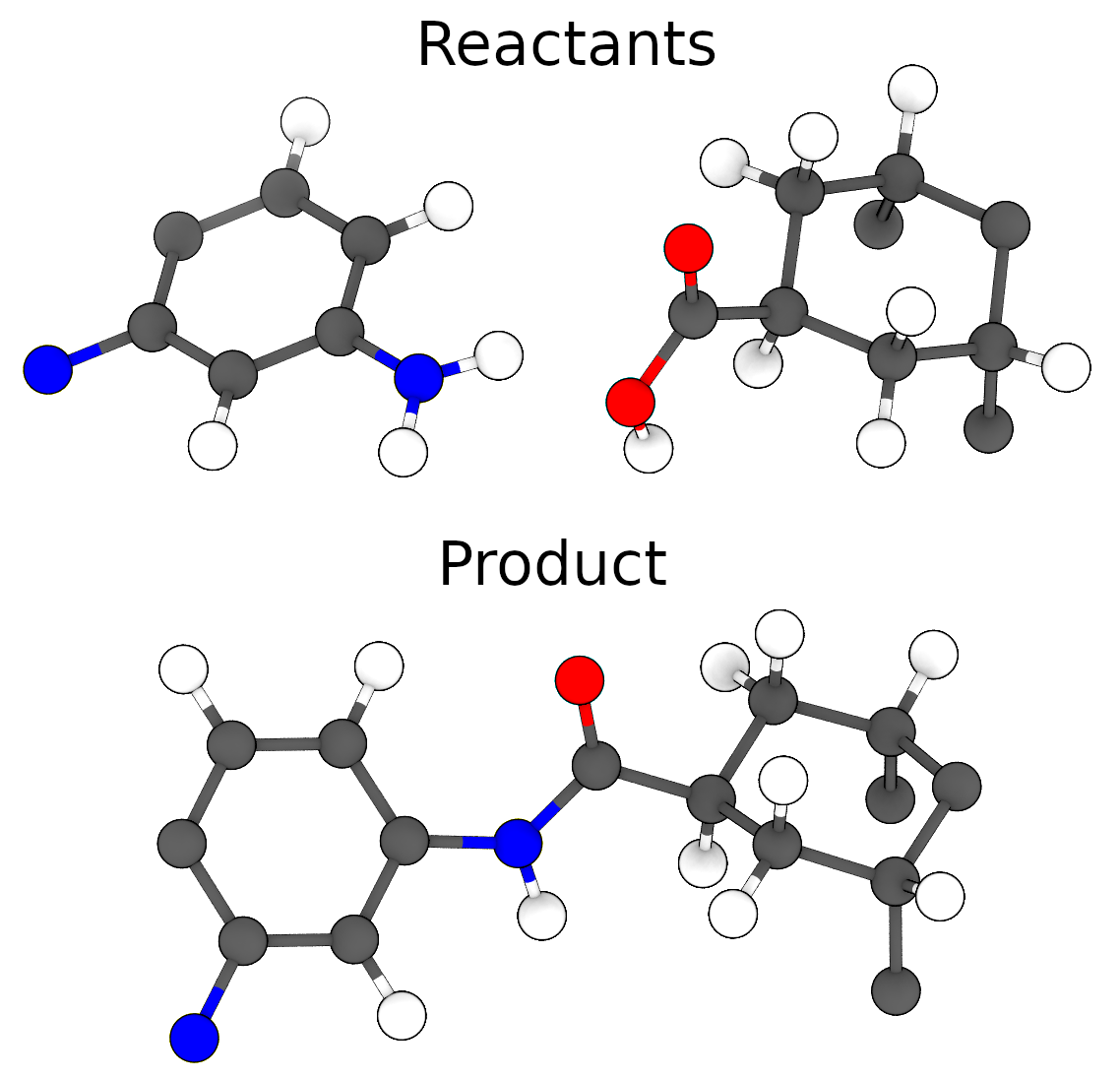}
    \end{minipage} \hfill
    \begin{minipage}{0.58\textwidth}
        \centering
        \includegraphics[width=1.0\textwidth]{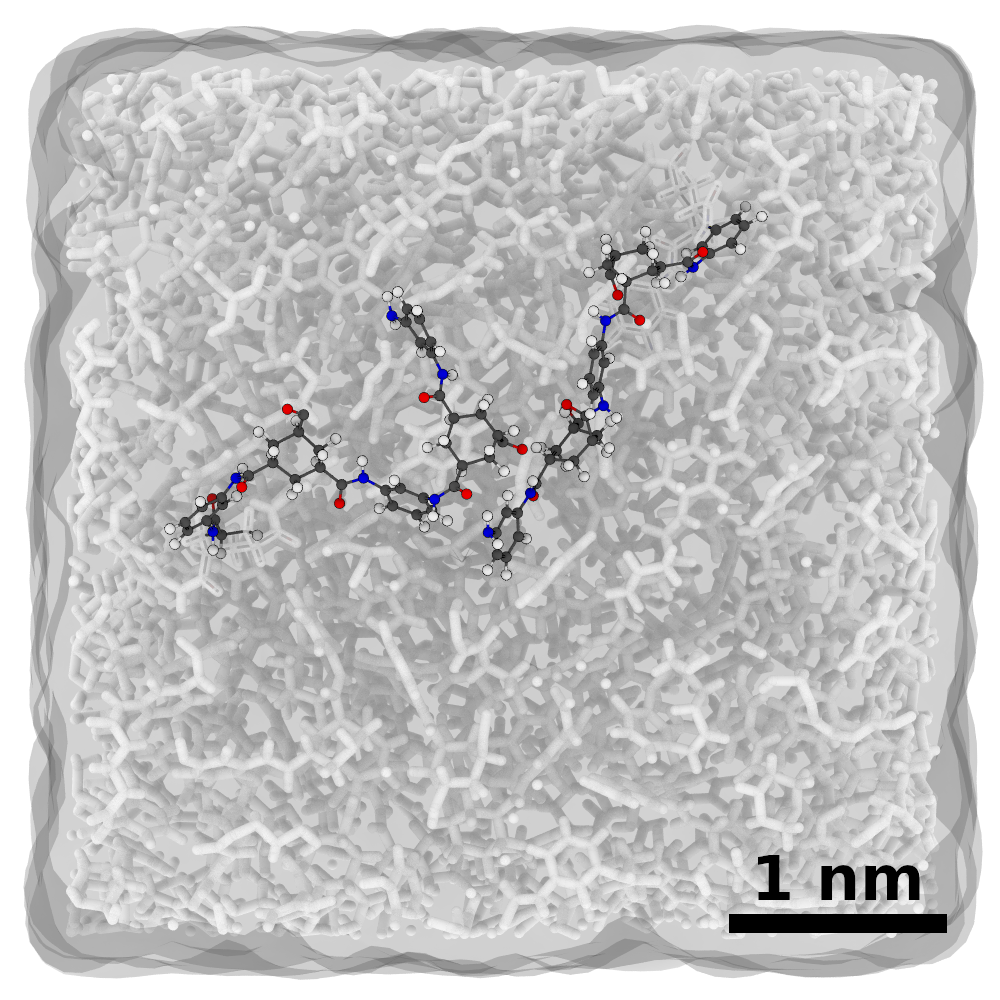}
    \end{minipage}
    \caption{The templates of the reactants and the product are molecular fragments of the reacting molecules (left). Their size is minimized and accounts for intramolecular interactions that change during the reaction. The illustration depicts a cross-linked polymer structure (right) with an edge length of approximately 50 {\AA}. The highlighted molecular strands represent the crosslinking.}
    \label{fig:system}
\end{figure}

To validate the algorithm-generated templates, a mixture of HT and MPD was prepared at an exact stoichiometric ratio, achieving an initial density of 1.26\,\si{g \per cm^3}. The crosslinking process achieved a degree of crosslinking (\(\eta\)) of 93\%. The final polymer structure exhibited a density of 1.20\,\si{g \per cm^3}, close to the the experimental value of 1.40\,\si{g \per cm^3} \cite{Arthur1989}. This deviation can be attributed to simulation constraints such as the limited box size (\(l_{\text{box}} = 48\)\,{\AA}) and the simplified reaction conditions. Nonetheless, the results confirm that the algorithm enables accurate modeling of polymer structures starting directly from SMILES strings, as depicted in Figure~\ref{fig:system}. \\

\begin{figure}[htb!]
    \centering
    \begin{minipage}{0.33\textwidth}
        \centering
        \includegraphics[width=1.0\textwidth]{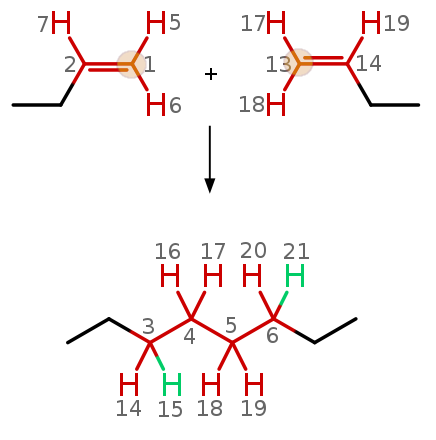}
    \end{minipage} \hfill
    \begin{minipage}{0.63\textwidth}
        \centering
        \includegraphics[width=1.0\textwidth]{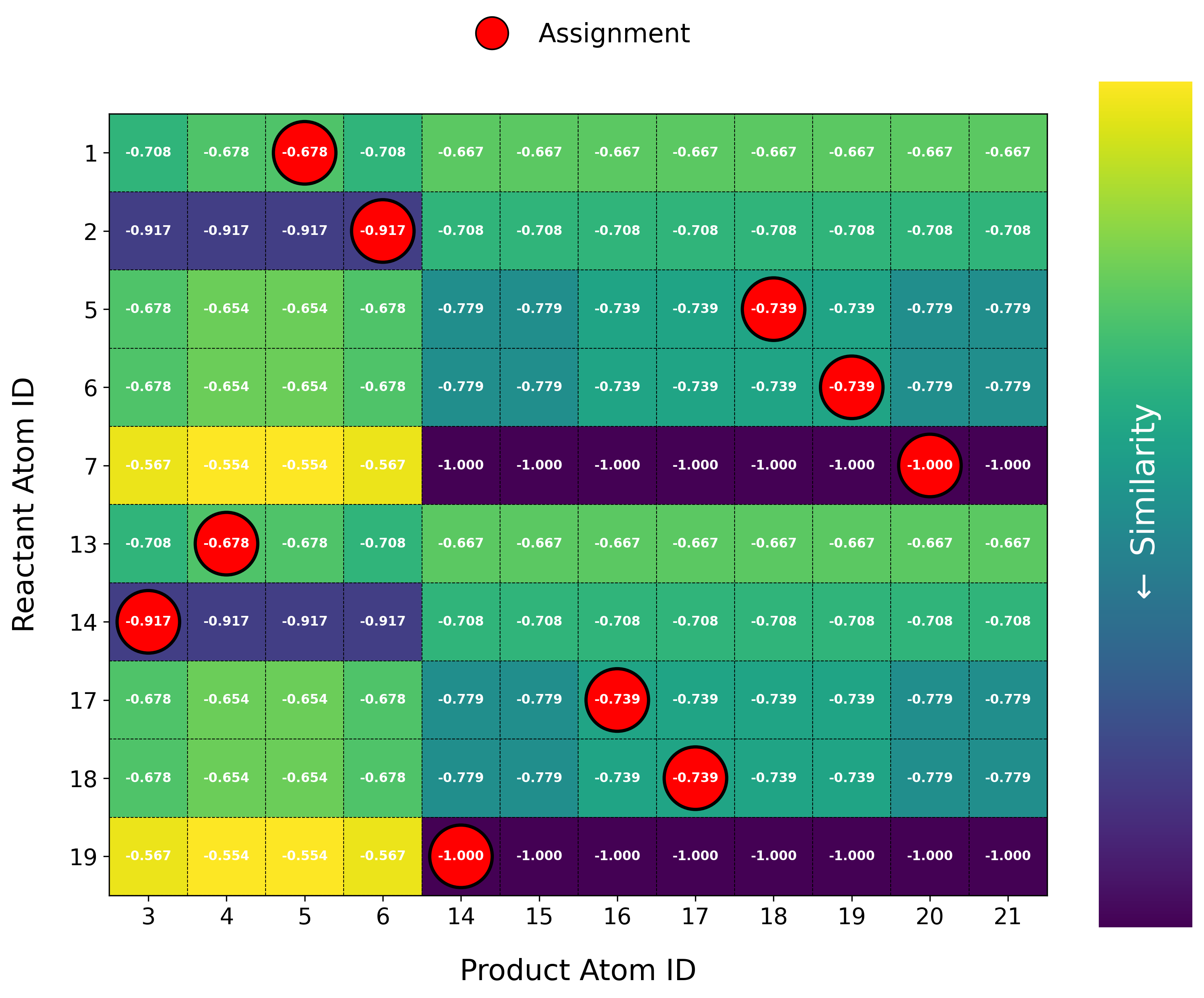}
    \end{minipage}
    \caption{The left panel illustrates the chain polymerization reaction type, where the ethylene group (black) is identified as the only conserved domain. The similarity matrix reveals ambiguities for several atoms based on the calculated similarity scores. The correct atom mapping was achieved through path-dependent and neighborhood-based swapping, as shown by the atomic labels in the chemical structures. The green hydrogen atoms are introduced during the alchemical reduction process, and the reaction-initiating atoms are highlighted in orange.}
    \label{fig:chain_condensation}
\end{figure}

The reduction of 1-butene to octane, as an example of an alchemical chain reaction, demonstrates the algorithm's ability to handle symmetrical reactants and complex atom mapping scenarios. This reaction involves the formation of a cross-link between the double bonds of two butene molecules, resulting in a saturated octane product. As shown in Figure~\ref{fig:chain_condensation}, 14 out of the 26 total nodes were identified as conserved domains using sub-graph isomorphism search. These conserved regions primarily correspond to the ethylene group, which remains unaltered during the reaction and is highlighted in black.

Using similarity scores, the algorithm initially mapped the nodes via linear sum assignment. However, ambiguities arose due to the indistinguishability of equivalent atoms between the two identical reactants (e.g., Node 2 and Node 14). This challenge was addressed through an iterative refinement process, where graph paths originating from isomorphic anchors were analyzed, allowing for the swapping of indistinguishable atoms and ensuring consistent mapping.

The mapping refinement also accounted for newly formed hydrogen atoms, which exhibited high similarity to existing hydrogen atoms. Neighborhood analysis proved crucial in resolving these ambiguities by evaluating structural and connectivity changes within the molecular graphs. Specifically, the assignment of atoms 5 and 6 to atoms 16 and 17, as well as atom 7 to either 14 or 15, could only be achieved through neighborhood analysis. This allowed for the clear identification of atoms 15 and 21 as those created during the reaction.

The final mapping is shown in Figure~\ref{fig:chain_condensation} (left), where atomic labels confirm the successful resolution of the potential misalignments. The reaction-initiating atoms highlighted in orange and the hydrogen additions in green. The bond rearrangements at the double bonds, were accurately identified. The final reaction template, comprising all 24 atoms, was derived as all atoms of the molecules are directly involved in the reaction and is available in the supplementary materials.

While in general the similarity scoring parameters (\(\alpha\), \(\beta\) and \(\gamma\)) can be used with the proposed default values, their adjustment may be necessary for reactions with fundamentally different chemistry or higher structural complexity. The observed variability in conserved regions also underscores the role of the force fields in accurately defining atom types and local environments. In reactions where specific atom types or bonding interactions are not well-defined by the chosen force field, the algorithm may face challenges in accurately identifying conserved domains and reaction sites. Consequently, the choice of force field becomes a crucial factor in determining the success of the algorithm with default parameters. Additionally, the current method may exhibit limitations in systems involving significant symmetry, conformational flexibility, or alchemical atom creation, where the alignment of molecular paths becomes more challenging.

\section*{Conclusions}

In this work, a graph-based algorithm for the automated generation of reaction templates for the \texttt{LAMMPS} package REACTION was successfully developed and demonstrated. Leveraging graph-theoretical principles and sub-graph isomorphism techniques, this method eliminates the labor-intensive manual processes traditionally required in reaction template preparation. By solely relying on \texttt{LAMMPS} input data, our approach ensures scalability, consistency, and adaptability across diverse reaction types.

The algorithm was validated on multiple case studies, including poly-addition, poly-condensation, and chain-polymerization reactions. These examples showcased the ability to accurately map conserved domains, identify reaction-initiating sites, and handle ambiguous molecular alignments through iterative refinement processes. Notably, the tool demonstrated high efficiency in reducing template size while preserving crucial molecular interactions, providing streamlined outputs for large-scale simulations.

While the presented framework focused on polymerization reactions, it has the potential to be extended to any organic reaction. Its flexibility is evident in the adjustable similarity scoring parameters, which can be tailored to capture the nuances of different chemical systems. Moreover, the algorithm’s graph-based design inherently supports integration with machine learning pipelines for high-throughput material discovery and structural data generation \cite{Baudis2021, Hayashi2022}.

Despite its strengths, certain limitations warrant further investigation. For instance, the accuracy of reaction site identification is influenced by the quality of the force fields employed, particularly regarding their ability to distinguish interactions between different atom types based on local atomic environments. Addressing this dependency through adaptive or transferable scoring mechanisms could enhance the algorithm's robustness. Additionally, improving its capability to handle highly symmetrical or complex molecules could broaden its application scope.

Looking ahead, this tool opens new opportunities for automating reaction modeling in computational chemistry and materials science. Future work may explore coupling this algorithm with quantum mechanical calculations or extending it to simulate reactive systems in mixed organic-inorganic environments. Furthermore, integrating this method with cloud-based platforms could facilitate collaborative, large-scale simulation workflows.

In conclusion, the algorithm represents a significant step toward automating and democratizing the preparation of reaction templates for molecular simulations. By bridging the gap between molecular mechanics and high-throughput computational workflows, it has the potential to accelerate innovation in fields ranging from polymer science to drug discovery.

\section*{Data Availability Statement}

All data and scripts supporting the findings of this study are openly available at the \texttt{Templater} repository via \url{https://collaborating.tuhh.de/m-29/software/templater}. This includes the full source code, datasets, and relevant documentation. The version of the data related to this study is available on the \texttt{published} branch.

\section*{Acknowledgements}

This research received funding from Grant No. 525597740 provided by the Deutsche Forschungsgemeinschaft (DFG, German Research Foundation).

\section*{Conflict of interest}

The authors declare that they have no competing interests to declare.

\bibliographystyle{ieeetr} 
\bibliography{ref.bib}

\end{document}